\documentclass[prb,reprint,nofootinbib,nobibnotes,showpacs]{revtex4-1}
\usepackage{graphicx}
\usepackage{xcolor}
\usepackage{amsmath}

\begin{document}

\title{Density-functional Monte-Carlo simulation of CuZn order-disorder transition}

\date{\today}
\author{S. N. Khan}
\email{khansn@ornl.gov}
\author{Markus Eisenbach}
\affiliation{Materials Science and Technology Division, Oak Ridge National Laboratory, Oak Ridge, TN 37831-6114}

\begin{abstract}
We perform a Wang-Landau Monte Carlo simulation of a Cu$_{0.5}$Zn$_{0.5}$ order-disorder transition using 250 atoms and pairwise atom swaps inside a $5\times5\times5$ BCC supercell. Each time step uses energies calculated from density functional theory (DFT) via the all-electron Korringa-Kohn-Rostoker method and self-consistent potentials. Here we find CuZn undergoes a transition from a disordered A2 to an ordered B2 structure, as observed in experiment. Our calculated transition temperature is near 870 K, comparing favorably to the known experimental peak at 750 K. We also plot the entropy, temperature, specific-heat, and short-range order as a function of internal energy.
\end{abstract}

\thanks{This manuscript has been authored by UT-Battelle, LLC, under
  Contract No. DE-AC05-00OR22725 with the U.S. Department of
  Energy. The United States Government retains and the publisher, by
  accepting the article for publication, acknowledges that the United
  States Government retains a non-exclusive, paid-up, irrevocable,
  world-wide license to publish or reproduce the published form of
  this manuscript, or allow others to do so, for United States
  Government purposes.}

\maketitle

CuZn is among the class of Hume-Rothery alloys.\cite{HumeRothery} Here the atomic size and crystal structure of the base metals are similar. As a result, electronic effects dominant phase stability mechanisms.\cite{PhysRevLett.67.1779} A key parameter is the electron-per-atom ratio (and/or chemical potential). The $e/a$ ratio determines the Fermi surface and concomitant nesting mechanisms and energy pseudogaps that can drive phase stability.\cite{0953-8984-17-50-013} For low $e/a$ on the Cu-rich side of the phase diagram, an A1 (FCC) solid solution is stable. On the Zn-rich side there are a series of complex, partially ordered phases. Of interest to us is Cu$_{0.5}$Zn$_{0.5}$, where the BCC structure is stable as a result of the Fermi surface crossing the Brillouin Zone boundary.\cite{0305-4608-1-2-303} Here an order-disorder transition occurs at T$_{c}$=750 K, \cite{Kowalski93, Miodownik} taking the system from a disordered A2 (BCC) phase to an ordered B2 phase (CsCl structure). We have sought to characterize this transition by using first-principles DFT and direct ensemble averaging through Monte Carlo simulation.

In order to reduce computational cost, methods for evaluating phase diagrams for binary alloys have either used model Hamiltonians or mean-field techniques. In the model Hamiltonian approach it is typical to expand the energy of an alloy configuration using a sum of nearest neighbor clusters weighted by undetermined coefficients.\cite{2002MSMSE..10..685Z} Formally this sets the energy $E=\sum_{\kappa}E_{\kappa}\prod_{i\in \kappa}\sigma_{i}$ where $\kappa$ stands for a cluster and $\sigma_{i}$ is a spin-like variable representing the occupancy on a site $i$.\cite{0034-4885-71-4-046501} The cluster coefficients $E_{\kappa}$ are chosen by comparing to DFT energies at a specified set of intermetallics. The geometry of the clusters can be identified a priori for a given structure and the energy of a given configuration rapidly evaluated once the coefficients are known. This enables Monte Carlo simulations to predict the phase diagram. This method has been applied to $\alpha$-brass (disordered FCC), finding a number of long-period superstructures at low temperatures ($\sim$20 K).\cite{PhysRevB.63.094204}
 
The other technique that has been employed is to perform a Landau-like perturbation theory from a mean-field disordered state.\cite{PhysRevLett.50.374} Here the perturbative order parameters are infinitesimal and independent concentration waves. A concentration wave imposes a variation on the uniform disordered medium by imposing a partial ordering along some direction. This could be, for example, a marginal increase (decrease) in the concentration of Cu for every even (odd) plane along (100). Within the KKR-CPA framework, the sign of the resulting free-energy change may be calculated.\cite{PhysRevB.50.1450} If negative at some critical temperature T$_{c}$, it predicts the disordered phase is unstable to this concentration variation. The incipient, unstable concentration wave can be used to anticipate a phase transition to a closely corresponding intermetallic. As an electronic theory it is capable of incorporating Fermi surface mechanisms. Using this method, Turchi et al.\cite{PhysRevLett.67.1779} found $\alpha$-brass to be the stable high-T phase of Cu$_{0.5}$Zn$_{0.5}$. On including phonon contributions they find the A2 ($\beta$) phase is more stable at high-T and that it transitions to B2 below T$_{c}$=700 K. We repeated this calculation using codes made available to us by J. Staunton at the Univ. of Warwick. On neglecting so-called charge-transfer effects we found a transition to an ordered B2 phase at 925 K. A correction to the mean-field medium due to Onsager that preserves certain sum rules on the short-range order parameter reduces this temperature to 615 K. We did not attempt to include phonon contributions as the structure was fixed to BCC ($\beta$-) brass.
 
What has not been done until now is to attempt a direct ensemble average using first-principles DFT for each configurational energy. This has been considered computationally infeasible, especially for cell sizes that begin to approach the thermodynamic limit. We show in this study that such a simulation is within reach and produces sensible results. We have performed a direct, Wang-Landau Monte Carlo simulation of a 250 atom CuZn supercell using first-principles DFT. No use of model Hamiltonians or fitting or expansions about a mean-field medium are performed. The total sample space is very large, consisting of $(250\text{ choose 125)}=10^{74}$ configurations. Here we sampled over 600,000 configurational energies, a calculation of unprecedented scale and close to our computational limit. Nevertheless we obtain a smooth density of states. Our calculation showcases the accuracy and limitations of first-principles DFT using ensemble averaging. It also serves as a benchmark for simulations that use cluster expansions and model Hamiltonians fit to DFT data. We find CuZn on a BCC lattice undergoes the predicted second-order transition, but at T$_{c}$=870 K. These calculations also demonstrate the possibility of direct calculation for other alloys, including multicomponent high-entropy alloys of recent interest \cite{doi:10.1080/21663831.2014.912690}.

Our Monte Carlo sampling is based on the Wang-Landau technique,\cite{PhysRevE.64.056101} a so-called flat histogram method. Such a method seeks to sample an energy window so that a Monte Carlo walker makes nearly equal visits to each energy bin. If configurations are selected randomly then this requires the probability to visit be weighted by $1/g(E)$ for density of states $g(E)$. In practice, a Wang-Landau run begins with guess density of states $g_{\text{app.}}(E)=1$. A random walker then makes moves in configurational space. Moves from an energy $E_{1}$ to $E_{2}$ are accepted with probability $p=\text{min}\{1,g_{\text{app.}}(E_{1})/g_{\text{app.}}(E_{2})\}$. At the end of each move, the guess density of states at walker position $E$ is improved by increasing $g(E)\rightarrow f\,g(E)$ for some modification factor $f>1$. This continues to bias the walker to energies with lower density-of-states. As a result, the histogram $H(E)$ of walker visits flattens as the simulation proceeds. Once a certain flatness criterion is achieved, the modification factor is reduced and the histogram reset. The accuracy of the final density of states depends on the flatness criterion used and the final modification factor f. In our simulation we used modification factor $\log f=3.125\times10^{-4}$. The flatness criterion was $[\text{min }g(E)]/[\text{average }g(E)]>0.60$. An advantage of the Wang-Landau sampling technique is that the simulation may be run once and the temperature set a posteriori. This is true as long as the desired temperature is within the sampled energy window.

\begin{figure}
 \includegraphics[scale=0.48]{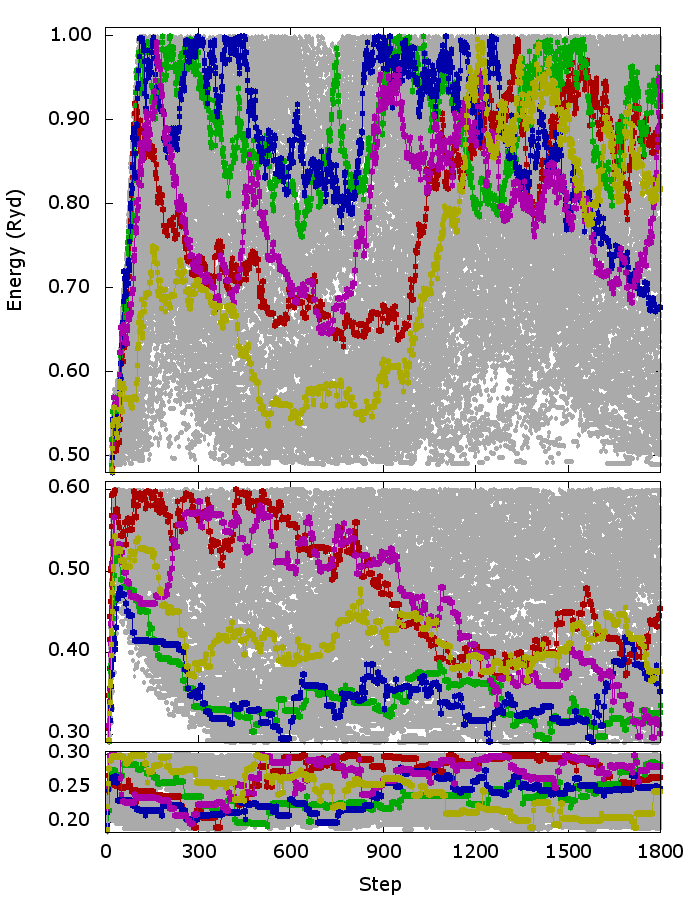}
\caption{(Color online) The energy trace of 125 Wang-Landau walkers for Monte Carlo runs in energy windows [0.5, 1.0], [0.3, 0.6], and [0.2, 0.3] respectively. Each window was performed as a separate run. Grey dots represents the energy of a walker at some time step. Colored lines are the explicit trace of five walkers. After a warm-up period of up to 300 steps, the walkers cover the whole energy range. All walkers were initialized in the B2 ground-state for each run.}
\label{fig:etrace}
\end{figure}

Our simulation cell is a $5\times5\times5$ lattice of conventional BCC cells (or 250 atomic sites). The lattice spacing is taken from the experimental high-T phase as $a_{0} = 5.58414$ Bohr \cite{doi:10.1143/JPSJ.33.1350} and is fixed for all temperatures. One of the benefits of studying Cu$_{0.5}$Zn$_{0.5}$ is that the lattice spacing undergoes minimal change through the transition. In the low-T phase the spacing increases to 5.5902 Bohr, or a change of 0.1\%. \cite{Rao} Half the sites are set to Cu and the other half Zn. In high entropy (A2) configurations the Cu and Zn atoms are randomly distributed. In the ground state (B2) the Cu atoms are at conventional cube corners and Zn at the body center (or vice-versa). 

To calculate energetics and perform Wang-Landau sampling, we modified the all-electron KKR code LSMS3 at Oak Ridge National Lab.\cite{LSMS3} The LSMS3 code solves the Kohn Sham equations of density functional theory
using a real space implementation of the multiple scattering formalism \cite{Korringa1947,Kohn1954,Ebert2011}.
The  code achieves linear scaling of the computation effort for the number of atomic sites
by limiting the environment of each atom that will contribute to the calculations of the Green's function
at this atomic site. \cite{Wang1995}. The computational efficiency of the LSMS approach for large
supercells allows the direct use of constrained density functional energy calculations
inside classical Monte-Carlo simulations to calculate thermodynamic properties of materials
from first principles on modern supercomputers.\cite{Eisenbach2009a} The Wang-Landau implementation of the LSMS3 was originally designed for the thermodynamics of Fe or Ni spins. Here we modified the code to convert spin degrees of freedom to site occupancy variables $\{\xi_{i}\}$. $\xi_{i}=0$ (1) to indicate the presence of a Cu (Zn) atom at site $i$. 

Our move type is point-to-point atom swaps of unlike atoms. Small steps improve the acceptance ratio. This is especially helpful as the ground-state is approached. A steep density of states curve leads to significant slowdown in the Monte Carlo sampling due to large rejection rate. Atomic potentials are taken as spherical and total energies are calculated within the muffin-tin approximation.\cite{PhysRevB.9.3985} The energy includes the nuclear attraction, Coloumb repulsion, and exchange-correlation effects using the local-density functional parameterized by von Barth-Hedin\cite{vonBarth}. Approximately 30 iterations are required at each Monte Carlo step to achieve electronic self-consistency. This reduces the number of total Monte Carlo steps possible by the same factor. Other KKR details include basis cutoff LMAX=3 and LSMS local interaction zone of $8.5a_{0}$. These are typical parameters for metals within KKR. Note that the only connection between first-principles DFT and the Wang-Landau simulation is the total energy provided at each time step. The code was validated prior to simulation by ensuring energetics are invariant to serial vs. supercomputing runs and also invariant across symmetric configurations. We further confirmed that the B2 configurational energy (-3445.826295 Ryd/atom including core electron) was indeed lower than any other configuration simulated. Each energy is calculated to a precision of 10$^{-6}$ Ryd.

An initial attempt to Wang-Landau sample throughout the entire range of configurations had convergence issues. This was due to steep density of states near the ground-state. To mitigate this difficulty, we performed three separate Wang-Landau runs. One each in the energy windows from [0.2, 0.3]; [0.3, 0.6]; and [0.5, 1.0] Ryd. A restricted energy window limits the range of possible density-of-states and therefore improves acceptance ratios and reduces runtimes. All walkers were initialized to the ground-state configuration for each run. This reduced the warm-up time because moves generating moves toward higher density-of-states occur more often than the reverse. Using first-principles DFT on a 250 atom cell restricted our runs to just under 2000 Monte Carlo steps per walker. Nevertheless the resulting density of states curve is smooth as we used 125 walkers. Each walker consisted of 32 nodes and each node compromised 8 CPU cores and an Nvidia GPU. The complete configuration of each walker at each step was saved for post-processing purposes.

\begin{figure}
 \includegraphics[scale=0.50]{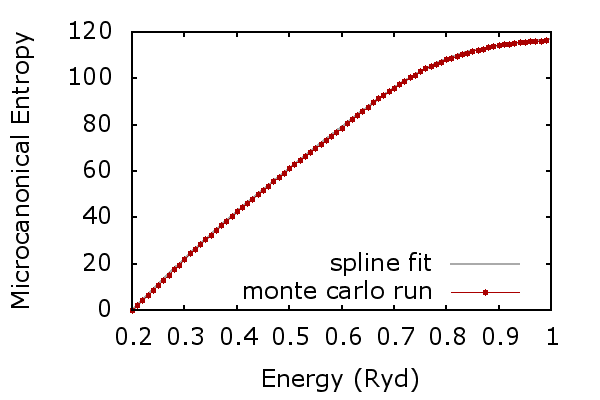}
\caption{(Color online) The sampled microcanonical entropy (red points) for 250 atom CuZn supercell and the corresponding cubic spline fit to data (solid gray). Each point represents a histogram bin from the Wang-Landau simulation. The microcanonical entropy is also the logarithm of the density of states. The entropy is shifted to zero at 0.2 Ryd, the lowest energy sampled.}
\label{fig:entropy}
\end{figure}

\begin{figure}
 \includegraphics[scale=0.50]{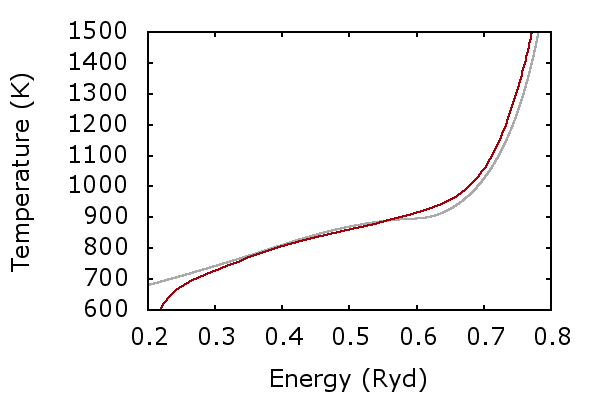}
\caption{(Color online) Temperature computed in the microcanonical (grey) and canonical ensembles (red). In the microcanonical ensemble we calculate $T=dU/dS$ from the cubic spline fit. In the canonical ensemble we calculate the Boltzmann average $\langle E\rangle=\sum_{i}E_{i}e^{-E_{i}/k_{B}T}/Z$ for partition function $Z$. The agreement between the two ensembles suggests approach to the thermodynamic limit. The Boltzmann average is incorrect below 700 K because the dominant term in the summation is below 0.2 Ryd.}
\label{fig:temp}
\end{figure}

In Fig.~\ref{fig:etrace} the energy trace from each energy window is presented. The walkers are less mobile in the lowest energy window. This is a result of a slowing down that occurs in this regime due to a steep density of states. However in all three windows by step 300 the walkers are sampling the entire range. Our main result is the logarithm of the density of states as presented in Fig.~\ref{fig:entropy}. Each point corresponds to an energy bin in the Wang-Landau simulation. The Wang-Landau method collects the density of states to within an arbitrary scale factor.  We fixed this factor by setting $\log g(0.2)=0$ at $E=0.2$ . We also scale the density of states in the other two windows to ensure continuity at $E=0.3$ and $E=0.6$. Our results may be interpreted in either the microcanonical or canonical ensemble. The extent to which the two approaches agree or disagree suggests how far we are from the thermodynamic limit.\cite{Binder19811655} In the microcanonical ensemble the internal energy is fixed and Fig.~\ref{fig:entropy} may be interpreted as the entropy up to an additive constant. Observables are calculated by taking the appropriate derivatives of the entropy curve. For this purpose we have fit the curve to a cubic spline. At $E=0$ the slope of the density of states will approach infinity (not visible). While in the totally disordered state at high energies (1 Ryd) the slope approaches zero. In between the slope is approximately constant over a large range of energies. This slope is in correspondance to the transition temperature T$_c$. At energies above 1 Ryd the slope becomes negative and this is only occurs at negative temperatures. In the canonical ensemble the temperature is fixed and instead the internal energy is calculated as a Boltzmann weighted average. The main disagreement we find between the two ensembles arises in shape and precise location of the peak in our specific heat curve. 

\begin{figure}
 \includegraphics[scale=0.50]{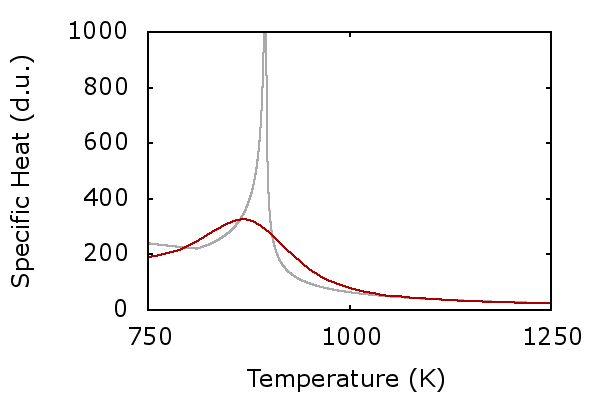}
\caption{(Color online) The specific heat calculated using $C_{V}=-\beta^{2}dU/d\beta$ from smooth spline fit of entropy (grey) and using $C_{V}=(\langle E^{2}\rangle-\langle E\rangle^{2})/(k_{B}^{2}T^{2})$ from Boltzmann weighted sums. Note that the calculation using a derivative is numerically less stable but capture the presence of a spike in specific heat near the phase transition. Using Boltzmann sums is numerically stable but approaches the correct limit with appreciable smoothing.}
\label{fig:spheat}
\end{figure}

\begin{figure}
 \includegraphics[scale=0.42]{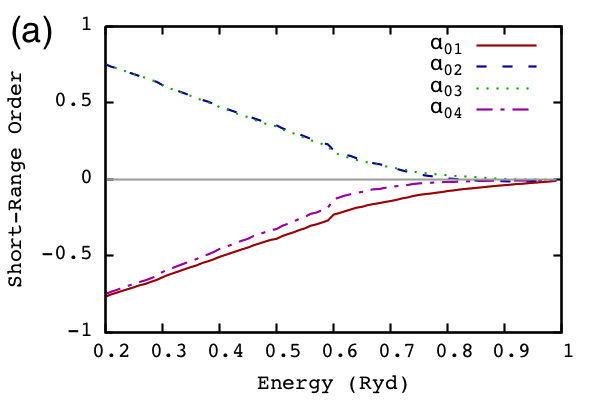}
  \includegraphics[scale=0.42]{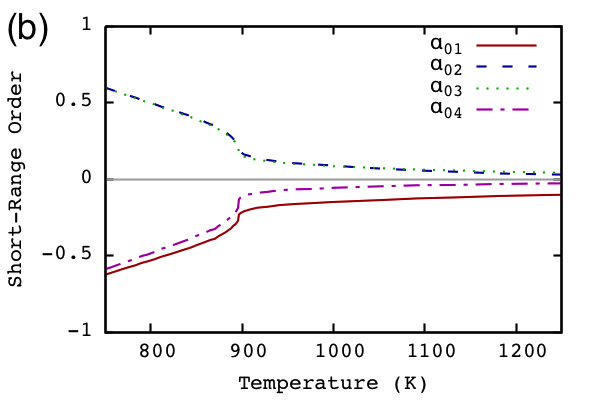}
\caption{(Color online) The short-range order parameters $c(1-c)\alpha_{ij}=\langle\xi_{i}\xi_{j}\rangle-\langle\xi_{i}\rangle\langle\xi_{j}\rangle$ for first four nearest neighbor shells containing 8, 6, 12, and 24 atoms respectively versus (a) energy and (b) temperature. As the ground-state (E=0) is approached, long-range order is established and all parameters go to either +1 or -1. Significant short-range order persists above the phase transition at E = 0.60.}
\label{fig:sro}
\end{figure}

Fig.~\ref{fig:temp} shows the relationship between the temperature and internal energy. In the microcanonical ensemble $T=\beta^{-1}=(dS/dU)^{-1}$ for $S=S(U)$ the entropy as given in Fig.~\ref{fig:entropy}. In the canonical ensemble the internal energy 
\[ U=\langle E\rangle=\int dEg(E)e^{-\beta E}/Z \]
 for partition function $Z=\int dE\,g(E)e^{-\beta E}$. For the system we consider, the density of states varies by many orders of magnitude. To prevent numeric overflow, the largest term in the sum is factored out. Numeric underflow remains but we ignore this since the dominant terms in the Boltzmann sum are usually included. However the Boltzmann weighted sum becomes invalid when the dominant term is outside the range [0.2, 1.0]. This is visible in the figure for T $<$ 700 K. Note that the curves calculated assuming two different ensembles otherwise overlay relatively well. This suggests the supercell shows signs of being in the thermodynamic limit. 
 
 The specific heat at constant volume is presented in Fig.~\ref{fig:spheat}. Again, this is computed for both ensembles. Because the transition is second-order we do not expect a latent heat of transformation. In the microcanonical we use 
 \[ C_{V}=-\beta^{2}\frac{dU}{d\beta}=-\beta^{2}(\frac{d^{2}S}{dU^{2}})^{-1} \] as calculated from the spline fit to the entropy. Here a sharp peak is evident at 895 K and results from $d\beta/dU$ approaching zero, as seen in Fig.~\ref{fig:entropy}. In the canonical ensemble we use 
 \[C_{V}=\frac{\langle E^{2}\rangle-\langle E\rangle^{2}}{k_{B}^{2}T{}^{2}} \]
and calculate $\langle E\rangle$ and $\langle E^{2}\rangle$ using Boltzmann weighted sums. The resulting curve peaks at 870 K and shows a smoother profile. This profile results from finite-size effects and would be a sharper peak for a large box.  At finite-size there are fluctuations in energy in the canonical ensemble that are not included in the microcanonical. In addition, performing numeric summations is more stable than taking numeric derivatives. Both computations are within reasonable agreement on $T_{c}$ however. 

There are two sources of error in the traditional Wang-Landau method: (1) An error from statistical sampling. It is clear from Fig.~\ref{fig:entropy} that our resulting density of states is quite smooth and much of this error has been eliminated. And, (2) An inherent bias because the method demands a minimum curvature in the resulting density of states. This has been examined by Brown \emph{et al.}\cite{PhysRevE.84.065702}, who calculate this minimum second-derivative as 
\begin{align*}
\frac{d^2}{dE^2} \log g(E) = \frac{\gamma}{\Delta E^2}
\end{align*}
where $\gamma = \log f$ is the modification factor and $\Delta E$ is the bin width. In our case $\gamma=0.004$ and $\Delta E = 0.01$ Ryd. We find the spline fit of our density of states has a second-derivative below this minimum only within the narrow range of energies [0.57, 0.606]. This is close to the critical temperature, which is to be expected as this is precisely where the curvature of the log density of states should be least.

In Fig.~\ref{fig:sro}, the Warren-Cowley short-range order parameters are presented. In the ground-state they approach -1 or 1. We see that short-range order is present and appreciable for temperatures above the phase transition. At 75$^{\circ}$ C above the calculated phase transition the short-range order magnitude is 0.19, 0.14, 0.13, and 0.09 for the first four shells respectively. In an neutron diffraction experiment on $\beta$-brass the short-range order at 75$^{\circ}$ C above the experimental transition using Zernike's theory is 0.18, 0.10, 0.07, and 0.05 respectively.\cite{Zernike1940565, PhysRev.130.1726} Note that an approximately linear relationship exists between the short-range order and configurational energy for E~$<$~0.7~Ryd. Focusing on the first shell, this suggests $E=\sum_i A\langle\xi_{i}\xi_{i+1}\rangle$ for some $A$. These first-principles DFT calculations lend support to the validity of model Hamiltonians based on nearest neighbor pair potentials. In Fig.~\ref{fig:sro}(b) a sudden increase in the short-range order is evident at the phase transition. In the thermodynamic limit this jump would be sharp and well-defined. 

In this paper we calculated the density-of-states for the CuZn binary alloy using a 250 atom unit cell and first-principles DFT to calculate energetics at each time step. We obtained a smooth density of states plot using over 600,000 samples. The lowest energy computed was a B2 ($\beta$-brass) ordering and the highest energies sampled showed total disorder. In Fig.~\ref{fig:spheat} a visible peak in the specific heat and sudden increase in atomic short-range order evident Fig.~\ref{fig:sro}b marks this transition. Using the canonical ensemble we find a critical temperature of 870 K. These results demonstrate the feasibility of performing direct first-principles Monte Carlo simulation without need for use of model Hamiltonians or mean-field expansions.

We acknowledge useful discussion with Ying Wai Li at Oak Ridge National Lab. This work was supported by the Materials Sciences \& Engineering Division of the Office of Basic Energy Sciences, U.S. Department of Energy. This research used resources of the Oak Ridge Leadership Computing Facility at ORNL, which is supported by the Office of Science of the U.S. Department of Energy under Contract No. DE-AC05-OOOR22725.

\bibliography{WangLandau_CuZn}

\end{document}